\documentclass[a4paper,12pt]{article}
\usepackage[dvips]{graphicx}
\usepackage{amssymb}
\usepackage{amsmath}

\begin{document}

\newcommand{\none}{$\mathcal{N}=1$ }
\newcommand{\be}{\begin{eqnarray}}
\newcommand{\ee}{\end{eqnarray}}
\numberwithin{equation}{section}

\newtheorem{theo}{Theorem}
\newtheorem{eg}{Example}

\title{Towards The Moduli Space of Extended Partial Isometries}
\author{Tewodros Amdeberhan$^1$\footnote{teddy@dimacs.rutgers.edu} and Arvind Ayyer$^2$\footnote{ayyer@physics.rutgers.edu} \\
$^1$DIMACS, Rutgers University \\
  96 Frelinghuysen Road \\
  Piscataway, NJ 08854\\[0.2in]
  $^2$Department of Physics \\
  Rutgers University\\
  136, Frelinghuysen Road \\
  Piscataway, NJ 08854}
\date{}

\maketitle

%\begin{abstract}
\centerline{\bf{Abstract}}
\noindent
Partial Isometries are important constructs that help give nontrivial solutions once a simple solution is known. We generalize this notion to Extended Partial Isometries and include operators which have right inverses but no left inverses (or vice versa). We find a large class of such operators and show the moduli space to contain the Hilbert Scheme of Points. Further, we apply this technique to find instanton solutions of a noncommutative \none supersymmetric gauge theory in six dimensions and show that this construction yields nontrivial solutions for other noncommutative gauge theories. The analysis is done in one complex dimension and the generalization of the result to higher dimensions is shown.

%\end{abstract}

\newpage

\section{Introduction}
One important idea used in solving problems in field theories is \cite{Harvey:2000jb} to find a transformation which is a symmetry of the equations of motion but not of the action. Acting by this transformation on a known, and typically simple, solution would yield other nontrivial solutions to the equations of motion. But these would be distinct solutions as far as the Lagrangian is concerned because there is no symmetry connecting the two actions.

In the context of noncommutative gauge theory, the objects used for solving such problems are called \emph{Partial Isometries}. These are operators $S$ which obey $S S^\dag S = S$. This implies that both $S S^\dag$ and $S^\dag S$ are projection operators. For our purposes it is enough that $S S^\dag$ is the identity operator. If $S^\dag S$ is also the identity, that would mean that $S$ is unitary. However this would not yield new solutions because unitary operations would be symmetries of the whole action. We generalize the concept of partial isometries to find new solutions of some noncommutative gauge theories. We center the discussion around a specific one and show that the technique works for some other noncommutative gauge theories.

The starting point is the study of solutions of a noncommutative gauge theory derived from a twisted \none supersymmetric model in six dimensions initiated in \cite{Iqbal:2003ds}. Their solutions correspond to instantons localized at the origin. We will show that this generalizes to instantons located anywhere. We demonstrate the workings of the mechanism in gory detail for the 1-instanton and the 2-instanton case and subsequently prove the existence of solutions corresponding to $n$-instantons (we explain the notation later) and show that this moduli space contains the Hilbert Scheme of Points as a subset. 

Because the idea of the construction was so simple, one could wonder why this was not discovered before. This is because there exists, in the literature, solutions of this kind known as the \emph{noncommutative ABS construction} \cite{Martinec:2001hh} described in a different way. 

The paper is organized as follows. In Section 2, we very briefly discuss the twisted scalar field theory and its noncommutative generalization. In section 3, we write down the ansatz and the simplest possible solution of this noncommutative field theory discussed in \cite{Iqbal:2003ds}. Section 4, which is the meat of the paper, contains explicit constructions of the extended partial isometries corresponding to 1-instanton and 2-instanton solutions and we show that these yield valid solutions to the noncommutative field theory discussed before. We then give the general prescription for extended partial isometries as well as an inductive proof. In Section 5, we prove that these extended partial isometries do solve the instanton equation of the noncommutative field theory discussed earlier. Section 6 contains a discussion of higher dimensional generalization in which we show that things generalize in a fairly straightforward manner. Finally, Section 7 shows that these solutions trivially yield solutions for some other well-known field theories.

\section{The Model and its Noncommutative solution}
The twisted \none $U(1)$ gauge theory under consideration was discovered in \cite{Baulieu:1997jx,Blau:1997pp,Bershadsky:1995qy,Acharya:1997gp,Donaldson:1996kp}. The bosonic part of the theory consists of $A$ - a $U(1)$ gauge field, $\phi$ - an untwisted complex scalar field, $\varphi$ - the twisted complex scalar field. 

The equations of motion are given by 
\be
F^{2,0}_A & = & \bar{\partial}^\dag_A \varphi \\
F^{1,1}_A \wedge k_0 \wedge k_0 + [\varphi, \bar{\varphi}] & = & l \phantom{i} k_0 \wedge k_0 \wedge k_0 \\
d_A \phi & = & 0
\ee
\noindent
where $k_0$ is the K\"ahler form on the manifold.

For the noncommutative solutions, here is the definition. Our manifold is $\mathbb{R}^6$ with the canonical commutators given by
\be
[x_m, x_n] & = & i \theta^{mn}, \qquad m,n = 1,...,6
\ee
\noindent
where $\theta$ is assumed to be of maximal rank. Without loss of generality, we can always bring $\theta$ to the canonical form involving $2 \times 2$ antisymmetric block matrices. 

Typically, it is more convenient to work with a linear combination of the gauge field and the coordinate. So define
\be
X^m = x^m + i \theta^{mn} A_n(x)
\ee

There are then natural complex objects we can work with. Define
\be
Z^{i} & = & \frac{1}{\sqrt{2 \theta_i}} (X^{2i-1} + i X^{2i})\\
Z^\dag_{i} & = & \frac{1}{\sqrt{2 \theta_i}} (X^{2i-1} - i X^{2i}), \qquad i=1,2,3
\ee
where $\theta_i$ is the (positive) $(1,2)$-th entry in the $i$th block.

For the gauge theory, (roughly speaking) we will simply replace covariant derivatives  by commutators with $Z$'s or $Z^\dag$'s and the K\"ahler form will be the standard symplectic form on $\mathbb{R}^6$.

Then the equations of motion are the following
\be
[ Z^i , Z^j ] + \epsilon^{ijk}[Z^\dag_k,\varphi] & = & 0 \qquad i,j = 1,2,3 \cr
[ Z^i , \phi] & = & 0 \qquad i=1,2,3 \cr
\sum_{i=1}^3 [ Z^i , Z^\dag_i ] + [ \varphi, \varphi^\dag ] & = & 3 
\ee

The solutions of these equations would yield instantons. We clarify the meaning of the operators. These objects are to be interpreted as operators acting on a Hilbert Space and these equations are to be interpreted as operator equations. The Hilbert space is defined as a vector space over $\mathbb{C}$ with basis given by $\{ |i,j,k \rangle | i,j,k \in \mathbb{Z}_+ \cup \{0\} \}$. This is also an orthonormal basis because of the inner product given by $\langle i,j,k | i',j',k'\rangle =\delta_{i,i'} \delta_{j,j'}\delta_{k,k'}$.

The vacuum solution is given by
\be
Z^i = a_i, Z^\dag_i = a^\dag_i, \varphi = 0, \phi = f \cdot 1
\ee 

where $f \in \mathbb{C}$ and $a_i, a^\dag_i$ are the standard harmonic oscillators raising and lowering operators acting as follows
\be
a_1 |n_1,n_2,n_3 \rangle & = & \sqrt{n_1} |n_1 - 1,n_2,n_3 \rangle \\
a^\dag_1 |n_1,n_2,n_3 \rangle & = & \sqrt{n_1+1} |n_1 + 1,n_2,n_3 \rangle
\ee
\noindent
and similarly for the other operators. By definition $a_1 |0,n_2,n_3 \rangle  = 0$ and similarly for the other $a_i$'s. They satisfy the commutation relation $[a_i, a^\dag_j] = \delta_{ij}$.

\section{The ansatz and one nontrivial solution}
Notice that if we set $\phi = f \cdot 1, \varphi =0$, we are left with solving a slightly simpler problem, namely,
\be
[ Z^i , Z^j ] & = & [Z^\dag_i,Z^\dag_j] = 0 \qquad i,j = 1,2,3 \cr
\sum_{i=1}^3 [ Z^i , Z^\dag_i ] & = & 3
\label{Zeqn}
\ee

The ansatz postulated for such problems was first constructed in \cite{Kraus:2001xt,Nekrasov:2002kc}
\be
Z^i & = & S a_i f(N) S^\dag \cr
Z^\dag_i & = & S f(N) a^\dag_i S^\dag
\label{ansatz}
\ee
\noindent
where $S$ is a partial isometry satisfying $S S^\dag = 1$ and $N = \sum a^\dag_i a_i$ is the number operator. $f(N)$ is the function to be determined from imposing the second equation in (\ref{Zeqn}). Note that the first two equations get automatically solved because of the ansatz. Therefore such an ansatz is not generically feasible when different directions (i.e. $Z$'s and $Z^\dag$'s) do not commute. 

In principle, it is entirely possible that we may find partial isometries which do not solve the above equations. That is to say, it is not possible to find a consistent function.

For simplicity, we'll initially solve this problem in one complex dimension.  The ansatz remains exactly the same in higher dimensions too. We label the states in the Hilbert space as $|i \rangle$ where $i \in \mathbb{Z}_+$ denotes the occupation number. The vacuum is $|0 \rangle$.

The simplest partial isometry one could postulate \cite{Witten:2000nz,Harvey:2000jt,Jatkar:2000ei}, sometimes called the `shift' isometry, is simply the following
\be
S^\dag |i \rangle & = & |i+m \rangle \cr
S |i \rangle & = & |i-m \rangle
\ee
\noindent
where $m>0$ is an integer. It is easy to see that $S S^\dag = 1$ but $S^\dag S = 1$ only in the subspace of the Hilbert space spanned by $\{ |m \rangle, |m+1 \rangle, \ldots \}$

Imposing this, we get
\be
Z |i \rangle & = & \sqrt{i+m} f(i+m) |i-1 \rangle \cr
Z^\dag |i \rangle & = & \sqrt{i+m+1} f(i+m+1) |i+1 \rangle
\ee

In one complex dimension, the first condition in (\ref{Zeqn}) is automatic so we just need to check the second one. Imposing it, one obtains
\be
(i+m+1) f^2(i+m+1) - (i+m) f^2(i+m) = 1
\ee

It is convenient to impose the boundary conditions $f(0) = f(1) = \ldots = f(m) = 0$ because the definition of $f(N)$ in this region is unimportant because of the nature of the ansatz. Setting $f(m) = 0$, we find a unique solution to the above equation, namely
\be
f(N) = \sqrt{1-\frac{m}{N}} \qquad N \geq m
\label{f1}
\ee

The way the solution is interpreted is the following. One looks at the image of the operator $S^\dag$. Generically, this will be a subspace generated by an ideal of the ring  of polynomials $\mathbb{C}[a^\dag]$ acting on the vacuum. One looks at the zero set of the ideal. That is where the instantons are said to be located. More precisely, the number of instantons is precisely the codimension of the ideal.

In this case, this is an ideal of the ring  of polynomials $\mathbb{C}[a^\dag]$ generated by monomials of the form $(a^\dag)^k$ where $k \geq m$. One looks at the zero set of the ideal. Thus, $m$ instantons are located at the origin in $\mathbb{C}$. Our convention is that an $n$-instanton denotes $n$ different locations for the instantons and not the number of instantons. So the shift isometry corresponds to a 1-instanton localized at the origin.

\section{Moduli space of Instantons}
We generalize this solution to two explicit cases, where the 1-instanton is located away from the origin and where 2-instantons are located away from the origin. It is here that we need to generalize the nature of the partial isometry. We call an operator $S$ an \emph{Extended Partial Isometry} if $\exists \, T \ni STS=S$. Then again, just as before, both $ST$ and $TS$ are projections and if both are the identity that is just the special case that $S$ is invertible.

Our ansatz is very close to the earlier one. We simply replace $S^\dag$ by $T$. We expect the function $f(N)$ to be unmodified for the first case simply because of the translation symmetry of $\mathbb{C}$. That is,
\be
Z^i & = & S a_i f(N) T \cr
Z^\dag_i & = & S f(N) a^\dag_i T
\label{ansatz2}
\ee

This might not seem satisfactory because now $Z^\dag \neq \left(Z\right)^\dag$ does not hold any more. (CHECK) One case where this might be useful is when we are dealing with complex spacetime. In that case, $X_i$'s are complex and then $Z_i$ is not required to be the conjugate of $Z^\dag_i$ any more.

%One way to interpret this is to say that non-commutativity is simplifying the relation between $Z$ and $Z^\dag$. Instead of requiring them to be Hermitian conjugates of each other, we require them to be ``inverses'' of each other in a loose sense. Namely,
\be
Z^{-1} & = & T^{-1} f(N) a^{-1} S^{-1} \\
& \sim & Z^\dag
\ee
\noindent
where $a^{-1}$, $T^{-1}$ and $S^{-1}$ are to be interpreted as the only possible inverses (left or right as the case may be) - $a^\dag$, $S$ and $T$ respectively. This is of course, very rough. One hopes to make this correspondence mathematically more rigorous.

We show that the space of Extended Partial Isometries contains the Hilbert Scheme of Points Hilb$_n$ in $\mathbb{C}[x]$. Recall that it is defined by 
\be
\textrm{Hilb}_n = \{ \textrm{Ideals} \, I \subset \mathbb{C}[x] \, | \,  dim_{\mathbb{C}}\mathbb{C}[x]/I = n \}
\ee
By definition, it is a smooth variety of complex dimension $2n$.

\subsection{1-instanton away from the origin}
For the first case, we want the ideal to be of the form $(x+p)^m$ where $p \in \mathbb{C}, m>0 \in \mathbb{Z}$. This forces $T$ to have the form
\be
T |i \rangle & = & (|1 \rangle + p | 0 \rangle )^{i+m} \cr
& = & \sum_{k=0}^{m+i} \binom{i+m}k p^k |i+m-k \rangle
\ee
\noindent
where the notation in the first equation is simply a mnemonic and the rigorous expression is the second one. Here, $\binom{i}k$ is the usual combinatorial factor. Then we need to determine the appropriate $S$. It is given by
\be
S |i \rangle & = & (|1 \rangle - p | 0 \rangle )^i | -m \rangle \cr
& = & \sum_{k=0}^{i} (-1)^k  \binom{i}k p^k |i-m-k \rangle
\ee

Note that there are no contributions to the summation if $k>i-m$. However, it will turn out to be important to include this range. This is because if $T$ acts after $S$, then we will also need terms from the region $k>i-m$.

The fact that $ST=I$ is just the simple combinatorial identity 
\be
\sum_{k=0}^m (-1)^k \binom{m}k  = 0
\ee

Note that the solution above reduces to the previous case in the limit $p \to 0$. The next thing is to solve the instanton equations with this ansatz. After reducing the upper limits, the equation which comes out is the following nontrivial equation.
\be
&&\sum_{j=0}^{i+1} \sum_{k=0}^{i-j+1} \sum_{l=0}^{i-j-k} \sum_{n=0}^{i-j-k-l} \Bigg[(-1)^{k+n} \phantom{i}  \binom{i-j-k+m-l}n \binom{i+m-j-k+1}l \cr
&\times& \binom{i+m}j \binom{i+m-j+1}k  \sqrt{i+m-j+1} \sqrt{i-j-k-l+m+1}   \cr
&\times& f(i+m-j+1) f(i-j-k-l+m+1)p^{j+k+l+n} |i-j-k-l-n \rangle\Bigg]\cr
&&- \sum_{j=0}^{i-1} \sum_{k=0}^{i-j-1} \sum_{l=0}^{i-j-k-1+m} \sum_{n=0}^{i-j-k-l} \Bigg[(-1)^{k+n} \phantom{i}  \binom{i-j-k+m-l}n \binom{i+m-j-1}k   \cr
&\times& \binom{i+m}j \binom{i+m-j-k-1}l  \sqrt{i+m-j} \sqrt{i-j-k-l+m}  \cr
&\times& f(i+m-j) f(i-j-k-l+m)p^{j+k+l+n}|i-j-k-l-n \rangle\Bigg] =|i \rangle
\label{f1eqn}
\ee

We prove that the only function that satisfies the above equation is the one given in (\ref{f1}) in the Section \ref{algo}.

\subsection{2-instantons, with one point as the origin} \label{2i1o}
As a warm up to the case of 2-instantons anywhere, we consider 2-instantons with one point the origin. Thus, the ideal is of the form $x^m (x+p)^n$ where, as before, $p \in \mathbb{C}, 0 < m,n \in \mathbb{Z}$. This gives
\be
T |i \rangle & = & (|1 \rangle + p | 0 \rangle )^{i+n} | m \rangle \cr
& = & \sum_{k=0}^{n+i} \binom{i+n}k p^k |i+m+n-k \rangle
\ee

Again, the tricky part is to postulate the action of $S$. Here it works exactly the same way as before
\be
S |i \rangle & = & (|1 \rangle - p | 0 \rangle )^{i-m} | -n \rangle \cr
& = & \sum_{k=0}^{i-m} (-1)^k \phantom{i}\binom{i-m}k p^k |i-m-n-k \rangle
\ee

The proof that $ST=I$ again follows exactly in the same way as before. Again, we could repeat the same exercise as before. Namely, calculate $Z| i \rangle$ and $Z^\dag | i \rangle$ and then evaluate $[Z, Z^\dag]| i\rangle$. Rather than doing it here, we'll show this result in the next section where we look at the general 2-instanton case. We just mention that the function involved is again very simple, namely
\be
f(N) = \sqrt{1 - \frac{m+n}{N}}
\label{f2}
\ee
\noindent
where we use initial conditions similar to the previous case. Namely, we set $f(0)=f(1)=\ldots=f(m+n)=0$. The solution corresponds to $m$ instantons at the origin and $n$ instantons at the point $-p$.

Again, let us look at special cases. In the limit $p \to 0$, this reduces to the $m+n$ instantons at the origin. Note that the function is also of the same form. Putting $m=0$ we get the previous solution. However, we do not get the shift isometry directly when we set $n=0$. We expect to see the shift isometry, and this is because with $n=0$, the extended partial isometry acts as follows
\be
T |i \rangle & = & (|1 \rangle + p | 0 \rangle )^{i} | m \rangle \cr
S |i \rangle & = & (|1 \rangle - p | 0 \rangle )^{i-m}
\ee
\noindent
and thus, the ideal is the same as that generated by $(a^\dag)^m$. The fact that these two ideals are the same leads to a highly nontrivial similarity transformation in the Hilbert space. The similarity matrix is a generalization of the so-called Stirling Numbers. The construction is given along with the proof in Appendix A.

However, this raises a question. Is it possible that our 1-instanton solution away from the origin also has a similarity transformation that takes it to the shift isometry? If so, that would jeopardize the whole program. Happily, that is not the case and it can be shown that there is no basis change that does the job. We can certainly see this from the two ideals since the ideal generated by $(a^\dag)^m$ and $(a^\dag -p)^m$ are not the same if $p \neq 0$.

\subsection{General 2-instanton case}
For the 2-instanton case, things get considerably more tricky. The binomial coefficients no longer have natural limits (that is, the summand does not go over exactly the part where the binomial coefficients are non-zero) and the proof that $ST=I$ is more involved.

The ideals are of the form $(x+p_1)^m (x+p_2)^n$ with both $0 \neq p_1,p_2 \in \mathbb{C}$ and $m,n \in \mathbb{Z}$. This time, postulating the extended partial isometry is highly nontrivial, especially for $S$. They are as follows
\be
T|i \rangle  & = & (|1 \rangle + p_1 |0 \rangle)^m (|1 \rangle + p_2 |0 \rangle)^{i+n} \cr
& = & \sum_{l=0}^m \binom{m}l p_1^l \sum_{k=0}^{n+i}  \binom{i+n}k  p_2^k |i+m+n-k-l \rangle \\
S|i \rangle  & = & (|1 \rangle + (p_1 -p_2) |0 \rangle )^{-m} (|1 \rangle -p_2 |0 \rangle)^i |-n \rangle \cr
& = & \sum_{l=0}^i (-1)^l \binom{i}l p_2^l  \cr
& & \times \sum_{k=0}^{i-m-n-l} (-1)^k \binom{m+k-1}k (p_1 - p_2)^k |i-m-n-k-l \rangle
\ee
\noindent
where, by definition, 
\be
\binom{-m}k = (-1)^k \binom{m+k-1}{k}
\ee
\noindent
and the upper limit of $k$ is set by the fact that $i-m-n-k-l \geq 0$. For convenience, we simplify the expression for the $S$ action.
\be
S|i \rangle  = \sum_{p=0}^{i-m-n} \sum_{q=0}^p (-1)^p \phantom{i} \binom{i-m-p+q}q \binom{m+p-q-1}{p-q} p_1^{p-q} p_2^q |i-m-n-p \rangle
\ee
\noindent
Then we see that $ST$ acts in the following way
\be
ST|i \rangle &=& \sum_{l=0}^m \sum_{k=0}^{n+i} \sum_{p=0}^{n+i-k-l} \sum_{q=0}^{p}(-1)^p  \binom{m}l \binom{i+n}k \binom{m+p-q-1}{p-q} \cr
& & \times \binom{i+n-k-l-p+q}q  p_1^{q+k} p_2^{p-q+l} |i -k-l-p \rangle
\label{st2}
\ee

The expression for the nontrivial part of (\ref{Zeqn}) is even more involved than before. We write it down for completeness.  Here is the expression.  It is a difference of two terms each having eight summation terms.
\be
& & \displaystyle \sum_{l=0}^{m} \sum_{k=0}^{i+n} \sum_{p=0}^{i+n-l-k+1} \sum_{q=0}^{p} \sum_{l2=0}^{m} \sum_{k2=0}^{i2+n} \sum_{p2=0}^{i2+n-l2-k2-1} \sum_{q2=0}^{p2} \Bigg[ \binom{m}l \binom{n+i}k \binom{m}{l2} \binom{i2+n}{k2}\cr
&\times& \displaystyle  \binom{i+n-k-l+1-p+q}q \binom{m+p-q-1}{p-q} \binom{m+p2-q2-1}{p2-q2} \cr
&\times& \binom{i2+n-k2-l2-1-p2+q2}{q2} p_1^{l+p-q+l2+p2-q2} p_2^{k+q+k2+q2}\cr
&\times& \displaystyle(-1)^{p+p2}  \sqrt{i+m+n-k-l+1} \sqrt{i2+m+n-k2-l2} f(i+m+n-k-l+1)\cr
&\times& \displaystyle  f(i2+m+n-k2-l2) \Bigg] |i-k-l-p-k2-l2-p2 \rangle \cr
&& \displaystyle - \sum_{l=0}^{m} \sum_{k=0}^{i+n} \sum_{p=0}^{i+n-l-k-1} \sum_{q=0}^{p} \sum_{l2=0}^{m} \sum_{k2=0}^{i3+n} \sum_{p2=0}^{i3+n-l2-k2+1} \sum_{q2=0}^{p2}\Bigg[  \binom{m}l \binom{n+i}k \binom{m}{l2} \binom{i3+n}{k2} \cr
&\times& \displaystyle  \binom{i+n-k-l-1-p+q}q \binom{m+p-q-1}{p-q} \binom{m+p2-q2-1}{p2-q2}\cr
&\times& \binom{i3+n-k2-l2+1-p2+q2}{q2} p_1^{l+p-q+l2+p2-q2} p_2^{k+q+k2+q2} \cr
&\times& \displaystyle (-1)^{p+p2}   \sqrt{i+m+n-k-l} \sqrt{i3+m+n-k2-l2+1} f(i+m+n-k-l)\cr
&\times& \displaystyle f(i3+m+n-k2-l2+1) \Bigg] |i-k-l-p-k2-l2-p2 \rangle = |i \rangle
\label{f2eqn}
\ee
\noindent
where $i2=i-k-l-p+1$ and $i3=i-k-l-p-1$. We know that there is a term proportional to $ |i \rangle$ when all the summands are zero. What we have to show is that the sum is zero for all terms with fixed $i-k-l-p-k2-l2-p2 < i$. As you can see things are getting pretty complicated. We prove the $n$-instanton identity in full generality in Section \ref{algo}.

As before, this corresponds to $m$ instantons at $-p_1$ and $n$ instantons at $-p_2$. The fact that $T$ is not symmetric in both the polynomials might seem strange. The expressions boil down to the previous cases when we set $m=0$ or $p_1=0$. If we set $m=0$, the solution reduces to the $1$-instanton away from the origin at $p_2$. $T$ reduces in an obvious way and to show that $S$ reduces, one notes that $\binom{p-q-1}{p-q}$ is zero unless $p=q$ in which case it is 1. In case of $p_1=0$, it again reduces to the same case but with $m+n$ instantons located at $p_2$. But they do not reduce to the previous case by setting $p_2=0$ or $n=0$. It would be an interesting exercise to analyze this limit and also to show that, as before, there is a similarity transformation taking this solution to the 1-instanton solution.

Notice also that we might as well have defined
\be
T|i \rangle  & = & (|1 \rangle + p_1 |0 \rangle)^{m+i} (|1 \rangle + p_2 |0 \rangle)^n \\
S|i \rangle  & = & (|1 \rangle + (p_2 -p_1) |0 \rangle )^{-n} (|1 \rangle -p_1 |0 \rangle)^i |-m \rangle 
\ee
\noindent
and then this choice would again define an extended partial isometry. And would also solve the instanton equation. To prove this, all we need to do is to interchange the role of $p_1$ and $p_2$ as well as $m$ and $n$.

This will be a separate solution if $p_1 \neq p_2$. If $p_1=p_2$, then we anyway reduce to a 1-instanton configuration. Thus, for a 2-instanton, we have two distinct ideals both giving rise to extended partial isometries and solving the instanton equation. We will see how this generalizes in the Section \ref{inst}.

\subsection{The Algorithm} \label{algo}
We describe the algorithm for finding generic extended partial isometries with $m$-instantons. The trick is to think of the basis consisting of $\{ |i \rangle | i\in \mathbb{Z}^{\geq 0} \}$ as the basis of the polynomial ring $\{ x^i | i\in \mathbb{Z}^{\geq 0} \}$. Then the actions of $S$ and $T$ described before are just the action on monomials in the ring.

For example, in the 1-instanton case,
\be
T(x^i) & = & (x+p)^{i+m} \\
S(x^i) & = & x^{-m} (x-p)^i 
\ee
\noindent
from which one can clearly see that $ST=I$. Thus, the action on an arbitrary polynomial $f(x)$ is given by
\be
T(f(x)) & = & (x+p)^m f(x+p) \\
S(f(x)) & = & x^{-m} f(x-p)
\ee
\noindent
where we remind the reader that these operations are to be seen as formal and make sense only when the exponent on $x$ is non-negative. $x^{-m}, m> 0$ is to be interpreted as 0. One way of looking at this is to consider the ring of rational functions and tensor these operators with the projection operator on the ring of polynomials. In the 2-instanton case, things are trickier but manageable. This time we get polynomials in the denominator
\be
T(x^i) & = & (x+p_1)^m (x+p_2)^{n+i} \\
S(x^i) & = & x^{-n}(x+p_1-p_2)^{-m} (x-p_2)^i
\ee
\noindent
where one is forced to break the symmetry between $(p_1,m)$ and $(p_2,n)$ in constructing $T$. Further, the $-m$ in the exponent is well-defined using the formula $\binom{-m}{k}=(-1)^k \binom{m+k-1}{k}$. $S$ is constructed by pure observation. Thus, on general polynomials,
\be
T(f(x)) & = & (x+p_1)^m (x+p_2)^n f(x+p_2) \\
S(f(x)) & = & x^{-n} (x+p_1-p_2)^{-m} f(x-p_2)
\ee

This gives us the prescription for higher $n$-instantons. Without further ado, here it is. Suppose our ideal of interest is generated by $(x+p_1)^{m_1} \ldots (x+p_n)^{m_n}$. Then the operators are
\be
T(f(x)) & = & (x+p_1)^{m_1} \ldots (x+p_n)^{m_n} f(x+p_n) \\
S(f(x)) & = & x^{-m_n} (x+p_1-p_n)^{-m_1} \ldots (x+p_{n-1}-p_n)^{-m_{n-1}} f(x-p_n)
\ee

%Of course, this is far from a proof. We conjecture that the corresponding combinatorial identities that arise at each level $n$ are true and moreover each give rise to solutions of (\ref{Zeqn}) in one complex dimension by the same ansatz as before. To prove these identities for all $n$ would require the aid of a computer unless there is a single unifying idea which would solve all of these identities in one stroke. 

We now prove the algorithm works by induction. Before that, we write down the action of $T |i \rangle$ and $S |i \rangle$ and deduce the identity to be proved for the $n$-instanton from the claim $ST |i \rangle =|i \rangle$. 

From the above construction, one can see that
\be
T | i \rangle & = & \sum_{k_1=0}^{m_1} \cdots \sum_{k_{n-1}=0}^{m_{n-1}} \sum_{k_n=0}^{i+m_n} \binom{m_1}{k_1} \cdots \binom{m_{n-1}}{k_{n-1}} \binom{i+m_n}{k_n} \cr
&\times & p_1^{k_1} \cdots p_n^{k_n} |i+\sum(m_j-k_j) \rangle \\
S | i \rangle & = & \sum_{l_n=0}^{i} \sum_{l_1+ \cdots + l_{n-1}=0}^{i-l_n}   \binom{-m_1}{l_1} \cdots \binom{-m_{n-1}}{l_{n-1}} \binom{i}{l_n} (-1)^{l_n}\cr
&\times &  (p_1-p_n)^{l_1} \cdots (p_{n-1}-p_n)^{l_{n-1}} p_n^{l_n} |i-\sum(m_j+l_j) \rangle
\ee
\noindent
where, for the purposes of this section, the sum denotes summing over the entire range of the variable being summed over. The upper limit on the $l_j$'s is established simply because of the condition $i-m_1-\cdots-m_n-l_1-\cdots-l_n>0$. Then the action of $ST$ after expanding each of the $(p_j-p_n)^{l_j}$'s is given by
\be
ST |i \rangle & = & \sum_{k_1=0}^{m_1} \cdots \sum_{k_{n-1}=0}^{m_{n-1}} \sum_{k_n=0}^{i+m_n}\sum_{l_n=0}^{i+\sum(m_j-k_j)} \sum_{l_1+ \cdots + l_{n-1}=0}^{i+\sum (m_j-k_j)-l_n} \sum_{q_1=0}^{l_1} \cdots \sum_{q_{n-1}=0}^{l_{n-1}} (-1)^{l_n+q_1+\cdots+q_{n-1}} \cr
& \times &  p_1^{k_1+l_1-q_1} \cdots p_{n-1}^{k_{n-1}+l_{n-1}-q_{n-1}} p_n^{k_n+l_n+q_1+\cdots+q_{n-1}} \binom{m_1}{k_1} \cdots \binom{m_{n-1}}{k_{n-1}} \binom{i+m_n}{k_n} \cr
& \times & \binom{l_1}{q_1} \cdots \binom{l_{n-1}}{q_{n-1}} \binom{-m_1}{l_1} \cdots \binom{-m_{n-1}}{l_{n-1}} \binom{i+\sum(m_j-k_j)}{l_n} |i-\sum(k_j+l_j) \rangle \nonumber
\ee

Since this is supposed to equal $|i \rangle$ for all possible values of $p_1,\ldots,p_n$, we fix the exponents of these variables. So, for $j=1,\ldots,n-1$, set $\alpha_j=k_j+l_j-q_j$ and set $\alpha_n=k_n+l_n+q_1+\cdots+q_{n-1}$. We use this to fix the values of $l_n$ and $q_1,\ldots,q_{n-1}$. Then the identity to be proved is
\be
A_n & := & \sum_{k_1=0}^{m_1} \cdots \sum_{k_{n-1}=0}^{m_{n-1}} \sum_{k_n=0}^{i+m_n} \sum_{l_1+ \cdots + l_{n-1}=0}^{\sum (\alpha_j-k_j)} (-1)^{k_n}  \binom{m_1}{k_1} \cdots \binom{m_{n-1}}{k_{n-1}} \binom{i+m_n}{k_n}  \cr
& \times & \binom{-m_1}{l_1} \cdots \binom{-m_{n-1}}{l_{n-1}} \binom{l_1}{k_1-\alpha_1+l_1} \cdots \binom{l_{n-1}}{k_{n-1}-\alpha_{n-1}+l_{n-1}} \cr
& \times & \binom{i+\sum(m_j-k_j)}{\sum (\alpha_j-k_j)-l_1-\cdots-l_{n-1}} = \delta_{\alpha_1,0} \cdots \delta_{\alpha_n,0}\label{An}
\ee

 To prove this by induction, we first prove this is true for $A_1=\delta_{\alpha_1,0}$. We then prove $A_n=\delta_{\alpha_n,0}A_{n-1}$. This would prove the identity for all $n$.

For $n=1$,
\be
A_1 & = & \sum_{k_1=0}^{i+m_1} (-1)^{k_1} \binom{i+m_1}{k_1} \binom{i+m_1-k_1}{\alpha_1-k_1} \cr
& = & \binom{i+m_1}{\alpha_1} \sum_{k_1=0}^{i+m_1} (-1)^{k_1} \binom{\alpha_1}{k_1}
\ee
But now, $\alpha_1 \leq i+m_1$ because $l_1 < i+m_1-k_1$ and $\alpha_1=l_1+k_1$. Thus the $k_1$-sum terminates at $\alpha_1$. But now,
\be
\sum_{k_1=0}^{\alpha_1} (-1)^{k_1} \binom{\alpha_1}{k_1} = \delta_{\alpha_1,0}
\ee
\noindent
which proves $A_1= \delta_{\alpha_1,0}$.

Now consider $A_n$. The way we are going to reduce it is by explicitly doing the $l_{n-1}$ sum and then the $k_{n-1}$ sum. So consider the terms involving $l_{n-1}$ in (\ref{An}). Using the definition of $\binom{-m}{k}$ given before and 
\be
& &\binom{m_{n-1}+l_{n-1}-1}{l_{n-1}}\binom{l_{n-1}}{k_{n-1}-\alpha_{n-1}+l_{n-1}}\cr
& = & \binom{m_{n-1}+l_{n-1}-1}{k_{n-1}-\alpha_{n-1}+l_{n-1}} \binom{m_{n-1}+\alpha_{n-1}-k_{n-1}-1}{\alpha_{n-1}-k_{n-1}}
\ee
\noindent
we have only two terms in the $l_{n-1}$-sum
\be
\sum_{l_{n-1}=0}^{\sum(\alpha_j-k_j)-l_1-\cdots-l_{n-2}} (-1)^{l_{n-1}} \binom{m_{n-1}+l_{n-1}-1}{k_{n-1}-\alpha_{n-1}+l_{n-1}}  \binom{i+\sum(m_j-k_j)}{\sum (\alpha_j-k_j)-l_1-\cdots-l_{n-1}} 
\ee

Since $\alpha_{n-1}-k_{n-1}>0$, the lower limit of the $l_{n-1}$-sum is forced to begin with $\alpha_{n-1}-k_{n-1}$. Make a change of variable $l'=\sum(\alpha_j-k_j)-l_1-\cdots-l_{n-1}$. Then we have
\be
& & (-1)^{\sum_{j=1}^{n-2}(\alpha_j-k_j-l_j)+\alpha_n-k_n} \sum_{l'=0}^{\sum_{j=1}^{n-2}(\alpha_j-k_j-l_j)+\alpha_n-k_n} (-1)^{l'} \cr
& \times & \binom{i+\sum(m_j-k_j)}{l'} \binom{m_{n-1}-1+\sum(\alpha_j-k_j)-l_1-\cdots-l_{n-2}-l'}{\sum_{j=1}^{n-2}(\alpha_j-k_j-l_j)+\alpha_n-k_n-l'}
\ee

This sum can now be solved using the Chu-Vandermonde identity. We finally get
\be
(-1)^{\alpha_{n-1}-k_{n-1}} \binom{i-\alpha_{n-1}+ \sum_{j=1}^{n-2}(m_j-k_j)+m_n-k_n}{\sum_{j=1}^{n-2}(\alpha_j-k_j-l_j)+\alpha_n-k_n}
\ee

Note that there is no $k_{n-1}$ factor in the expression. Therefore, there are only two binomial factors involving $k_{n-1}$ and there is hope that it can be solved. However, note that in (\ref{An}), the remaining $l_j$-sums are to be done before we do the $k_{n-1}$-sum. So we must find a way of switching the sums. To avoid cluttering the issue, let us call $a:=k_{n-1}$ and $b:=l_1+\cdots+l_{n-2}$. So we want to switch the $a,b$ sums in an expression of the form
\be
\sum_{a=0}^N \sum_{b=0}^{M-a} \nonumber
\ee

Now, there are two possibilities. If $N>M$, the $b$-sum forces the $a$-sum to go upto $M$ only. Then we can switch easily and
\be
\sum_{a=0}^N \sum_{b=0}^{M-a}=\sum_{b=0}^M \sum_{a=0}^{M-b} \nonumber
\ee

On the other hand, if $M>N$, we have to split the sum for $b$ upto $M-N$ and separately beyond it. Then, we get
\be
\sum_{a=0}^N \sum_{b=0}^{M-a}=\sum_{b=0}^{M-N} \sum_{a=0}^{N}+ \sum_{b=M-N}^M \sum_{a=0}^{M-b} \nonumber
\ee

Before we analyze both these cases, let us look at the summand. It is
\be
(-1)^{k_{n-1}} \binom{m_{n-1}}{k_{n-1}} \binom{m_{n-1}-1+\alpha_{n-1}-k_{n-1}}{\alpha_{n-1}-k_{n-1}}
\ee

The natural limits of the sum are from 0 to either of $m_{n-1}$ or $\alpha_{n-1}$. If the sum goes beyond any of these values, that is also fine because the coefficients will cut off the sum automatically. Then we can simplify the sum immediately using the Chu-Vandermonde convolution.

In the case $N>M$, the upper limit of the $k_{n-1}$-sum is $M-b=\sum \alpha_j-k_1-\cdots-k_{n-2}-k_n-l_1-\cdots-l_{n-2}$. But this is equal to $k_{n-1}+l_{n-1}+l_n$ by definition. Since $k_{n-1}+l_{n-1}>\alpha_{n-1}$ and $l_n>0$, the upper limit is definitely greater than or equal to $\alpha_{n-1}$.

In the other case, then $M>N$, the upper limit of the first term is $N=m_{n-1}$ which is exactly what we want. For the other case, it is once again $M-b$ and the same argument given in the previous paragraph goes through.  

Thus, in all the cases, we can solve the sum immediately to get $(-1)^{\alpha_{n-1}} \delta_{\alpha_{n-1},0}$.  For the first case, the upper limit is $M$. For the second case, the two terms add up to give the same upper limit.

One can now assemble the remaining terms to see that one gets exactly $\delta_{\alpha_{n-1},0} A_{n-1}$ where one has to simply exchange the notation for anything with an $n$ subscript and an $n-1$ subscript. This completes the induction argument and hence the proof.

\section{Proof of the Instanton Equation} \label{inst}
Having constructed and proved the existence of these extended partial isometries, one could wonder whether all of them yield solutions to the instanton equation $[Z,Z^\dag]=1$. In this section, we prove that they do and moreover, they all satisfy the above equation with the same function as for the 1-instanton and the 2-instanton and that the function is unique.

To begin with, we rewrite explicitly the forms of $S$ and $T$ for the $n$-instanton extended partial isometry.
\be
T | i \rangle & = & \sum_{k_1=0}^{m_1} \cdots \sum_{k_{n-1}=0}^{m_{n-1}} \sum_{k_n=0}^{i+m_n} \binom{m_1}{k_1} \cdots \binom{m_{n-1}}{k_{n-1}} \binom{i+m_n}{k_n} \cr
&\times & p_1^{k_1} \cdots p_n^{k_n} |i+\sum(m_j-k_j) \rangle \\
S | i \rangle & = & \sum_{l_n=0}^{i} \sum_{l_1+ \cdots + l_{n-1}=0}^{i-\sum m_j-l_n}   \binom{-m_1}{l_1} \cdots \binom{-m_{n-1}}{l_{n-1}} \binom{i}{l_n} (-1)^{l_n}\cr
&\times &  (p_1-p_n)^{l_1} \cdots (p_{n-1}-p_n)^{l_{n-1}} p_n^{l_n} |i-\sum(m_j+l_j) \rangle
\ee

Notice that the expression for $T$ has $n$ summations whereas $S$ has $2n-1$ when we expand all the $(p_j-p_n)^{l_j}$ terms. We start off simplifying this expression. First off, $S$ is given by
\be
S | i \rangle & = & \sum_{l_n=0}^{i} \sum_{l_1+ \cdots + l_{n-1}=0}^{i-l_n} \sum_{q_1=0}^{l_1} \cdots \sum_{q_{n-1}=0}^{l_{n-1}} (-1)^{l_1+\cdots+l_n+q_1+\cdots +q_{n-1}} \cr
& \times & \binom{m_1+l_1-1}{l_1} \cdots \binom{m_{n-1}+l_{n-1}-1}{l_{n-1}} \binom{i}{l_n} \binom{l_1}{q_1} \cdots \binom{l_{n-1}}{q_{n-1}} \cr
& \times & p_1^{l_1-q_1} \cdots p_{n-1}^{l_{n-1}-q_{n-1}} p_n^{l_n+q_1+\cdots q_{n-1}} |i - \sum (m_j+l_j) \rangle
\ee

Now change $l_n \to l_n'=l_1+ \cdots+l_n$. We then have
\be
S | i \rangle & = & \sum_{l_n'=0}^{i} (-1)^{l_n'} |i-l_n'-\sum m_j \rangle  \sum_{l_1+ \cdots + l_{n-1}=0}^{l_n'} \binom{m_1+l_1-1}{l_1} \cdots \binom{m_{n-1}+l_{n-1}-1}{l_{n-1}}\cr
& \times & \binom{i}{l_n'-l_1-\cdots l_{n-1}} \sum_{q_1=0}^{l_1} \cdots \sum_{q_{n-1}=0}^{l_{n-1}} (-1)^{q_1+\cdots +q_{n-1}} \cr
& \times &   \binom{l_1}{q_1} \cdots \binom{l_{n-1}}{q_{n-1}} p_1^{l_1-q_1} \cdots p_{n-1}^{l_{n-1}-q_{n-1}} p_n^{l_n'-l_1-\cdots-l_{n-1}+q_1+\cdots q_{n-1}}
\ee

Make the following change of variables: $l_j \to l_j'=l_j-q_j$ for $j=1,\ldots,n-1$. Then, the upper limit of the $q$ summations will change. Since $l_1,\ldots,l_{n-1}$ sums are treated together, this will force the $q_1,\ldots,q_{n-1}$ sums to also be treated together and we will have the following expression
\be
S | i \rangle & = & \sum_{l_n'=0}^{i} (-1)^{l_n'} |i-l_n'-\sum m_j \rangle \times \sum_{l_1'+ \cdots + l_{n-1}'=0}^{l_n'} p_1^{l_1'} \cdots p_{n-1}^{l_{n-1}'} p_n^{l_n'-l_1'-\cdots-l_{n-1}'} \cr
& \times & \sum_{q_1+ \cdots+ q_{n-1}=0}^{l_n'-l_1'-\cdots-l_{n-1}'} (-1)^{q_1+\cdots +q_{n-1}} \binom{m_1+l_1'+q_1-1}{l_1'+q_1} \cdots \binom{m_{n-1}+l_{n-1}'+q_{n-1}-1}{l_{n-1}'+q_{n-1}} \cr
& \times & \binom{i}{l_n'-l_1'-\cdots-l_{n-1}'-q_1-\cdots-q_{n-1}} \binom{l_1'+q_1}{q_1} \cdots \binom{l_{n-1}'+q_{n-1}}{q_{n-1}}
\ee

Now notice, that for all $j=1,\ldots,n-1$, we can replace $\binom{m_j+l_j'+q_j-1}{l_j'+q_j} \binom{l_j'+q_j}{q_j}$ by $\binom{m_j+l_j'+q_j-1}{q_j} \binom{m_j+l_j'-1}{l_j'}$. 

Then each of the $q_j$'s appear in exactly two binomial coefficients and they can be summed over using the Chu-Vandermonde identity. Notice that the limits are exactly what we want in the identity. Since the $q_j$'s appear together, one has to choose a particular ordering of the way the sum is done, but the final answer is independent of the order. So we finally have (after removing the primes)
\be
S | i \rangle & = & \sum_{l_n=0}^{i} \sum_{l_1+ \cdots + l_{n-1}=0}^{l_n} (-1)^{l_n} \binom{m_1+l_1-1}{l_1} \cdots \binom{m_{n-1}+l_{n-1}-1}{l_{n-1}}  \cr
& \times & \binom{i- \sum_{j=1}^{n-1} (m_j + l_j)}{l_n-l_1-\cdots-l_{n-1}} p_1^{l_1} \cdots p_{n-1}^{l_{n-1}} p_n^{l_n-l_1-\cdots-l_{n-1}} |i-l_n-\sum m_j \rangle
\ee

Using the ansatz in (\ref{ansatz2}), we construct $Z,Z^\dag$ with this formula for $S$ to get
\be
Z|i \rangle & = & \sum_{k_1=0}^{m_1} \cdots \sum_{k_{n-1}=0}^{m_{n-1}} \sum_{k_n=0}^{i+m_n} \sum_{l_n=0}^{i+\sum(m_j- k_j) -1} \sum_{l_1+ \cdots + l_{n-1}=0}^{l_n} (-1)^{l_n}   \cr
& \times & p_1^{k_1+l_1} \cdots p_{n-1}^{k_{n-1}+l_{n-1}} p_n^{k_n+l_n-l_1-\cdots-l_{n-1}} \binom{m_1}{k_1} \cdots \binom{m_{n-1}}{k_{n-1}} \binom{i+m_n}{k_n}   \cr
& \times & \binom{m_1+l_1-1}{l_1} \cdots \binom{m_{n-1}+l_{n-1}-1}{l_{n-1}} \binom{i+m_n-1-l_1-\cdots-l_{n-1}-\sum k_j}{l_n-l_1-\cdots-l_{n-1}} \cr
& \times & g\left(i+\sum(m_j-k_j)\right)|i-l_n-\sum k_j -1 \rangle \\
Z^\dag |i \rangle & = & \sum_{k_1=0}^{m_1} \cdots \sum_{k_{n-1}=0}^{m_{n-1}} \sum_{k_n=0}^{i+m_n} \sum_{l_n=0}^{i+\sum (m_j-k_j) +1} \sum_{l_1+ \cdots + l_{n-1}=0}^{l_n} (-1)^{l_n}   \cr
& \times & p_1^{k_1+l_1} \cdots p_{n-1}^{k_{n-1}+l_{n-1}} p_n^{k_n+l_n-l_1-\cdots-l_{n-1}} \binom{m_1}{k_1} \cdots \binom{m_{n-1}}{k_{n-1}} \binom{i+m_n}{k_n}   \cr
& \times & \binom{m_1+l_1-1}{l_1} \cdots \binom{m_{n-1}+l_{n-1}-1}{l_{n-1}} \binom{i+m_n+1-l_1-\cdots-l_{n-1}-\sum k_j}{l_n-l_1-\cdots-l_{n-1}} \cr
& \times & g\left(i+\sum(m_j-k_j)+1\right) |i-l_n-\sum k_j +1 \rangle
\ee
\noindent
where $g(N) = \sqrt{N} f(N)$. We now look at $Z Z^\dag$ and $Z^\dag Z$ separately with the following convention. We use the letter $k$ for the first $T$, $l$ for the first $S$, $r$ for the second $T$ and $s$ for the second $S$. Then we have
\be
Z Z^\dag |i \rangle & = & \sum_{k_1=0}^{m_1} \cdots \sum_{k_{n-1}=0}^{m_{n-1}} \sum_{k_n=0}^{i+m_n} \sum_{l_n=0}^{i+\sum (m_j-k_j) +1} \sum_{l_1+ \cdots + l_{n-1}=0}^{l_n} \sum_{r_1=0}^{m_1} \cdots \sum_{r_{n-1}=0}^{m_{n-1}} \sum_{r_n=0}^{i+m_n-l_n+1-\sum k_j} \cr
& \times & \sum_{s_n=0}^{i-l_n+\sum (m_j-k_j-r_j) } \sum_{s_1+ \cdots + s_{n-1}=0}^{s_n} (-1)^{l_n+s_n} \binom{m_1}{k_1} \cdots \binom{m_{n-1}}{k_{n-1}} \binom{i+m_n}{k_n} \cr
& \times &  \binom{m_1+l_1-1}{l_1} \cdots \binom{m_{n-1}+l_{n-1}-1}{l_{n-1}} \binom{m_1}{r_1} \cdots \binom{m_{n-1}}{r_{n-1}} \cr
& \times &   \binom{i+m_n+1-l_n-\sum k_j}{r_n} \binom{m_1+s_1-1}{s_1} \cdots \binom{m_{n-1}+s_{n-1}-1}{s_{n-1}} \cr
& \times &  \binom{i+m_n+1-l_1-\cdots-l_{n-1}-\sum k_j}{l_n-l_1-\cdots-l_{n-1}}\cr
& \times &  \binom{i+m_n-l_n-s_1-\cdots-s_{n-1}-\sum (k_j+r_j)}{s_n-s_1-\cdots-s_{n-1}} \cr
& \times & g\left(i+\sum(m_j-k_j)+1\right) g\left(i-l_n+\sum(m_j-k_j-r_j)+1\right) \cr
& \times & p_1^{k_1+l_1+r_1+s_1} \cdots p_{n-1}^{k_{n-1}+l_{n-1}+r_{n-1}+s_{n-1}} p_n^{k_n+l_n+r_n+s_n-l_1-\cdots -l_{n-1} -s_1-\cdots -s_{n-1}} \cr
& \times &  |i-l_n-s_n-\sum (k_j+r_j) \rangle
\ee

Since the identity is supposed to hold for any values of the $p_j$'s, we can fix their exponents. Let
\be
\alpha_1 & = & k_1+l_1+r_1+s_1 \cr
& \vdots & \cr
\alpha_{n-1} & = & k_{n-1}+l_{n-1}+r_{n-1}+s_{n-1} \cr
\alpha_n & = & k_n+l_n+r_n+s_n-l_1-\cdots -l_{n-1} -s_1-\cdots -s_{n-1}
\ee

We use this to fix the values of $r_1,\ldots, r_n$ in the usual way to get 
\be
& & p_1^{\alpha_1} \cdots p_n^{\alpha_n} |i- \sum \alpha_j \rangle \times \sum_{k_1=0}^{m_1} \cdots \sum_{k_{n-1}=0}^{m_{n-1}} \sum_{k_n=0}^{i+m_n} \sum_{l_n=0}^{i+\sum (m_j-k_j) +1} \sum_{l_1+ \cdots + l_{n-1}=0}^{l_n} \cr
& \times & \sum_{s_n=0}^{\sum (\alpha_j-k_j)-l_n } \sum_{s_1+ \cdots + s_{n-1}=0}^{s_n} (-1)^{l_n+s_n} \binom{m_1}{k_1} \cdots \binom{m_{n-1}}{k_{n-1}} \binom{i+m_n}{k_n}\cr
& \times &  \binom{m_1+l_1-1}{l_1} \cdots \binom{m_{n-1}+l_{n-1}-1}{l_{n-1}} \binom{m_1+s_1-1}{s_1} \cdots \binom{m_{n-1}+s_{n-1}-1}{s_{n-1}} \cr
& \times & \binom{m_1}{\alpha_1-k_1-l_1-s_1} \cdots \binom{m_{n-1}}{\alpha_{n-1}-k_{n-1}-l_{n-1}-s_{n-1}} \cr
& \times &   \binom{i+m_n+1-l_n-\sum k_j}{\alpha_n-k_n-l_n-s_n+l_1+\cdots+l_{n-1}+s_1+\cdots+s_{n-1}}  \cr
& \times &  \binom{i+m_n+1-l_1-\cdots-l_{n-1}-\sum k_j}{l_n-l_1-\cdots-l_{n-1}} \binom{i+m_n-\sum \alpha_j-s_1-\cdots-s_{n-1}}{s_n-s_1-\cdots-s_{n-1}} \cr
& \times & g\left(i+\sum(m_j-k_j)+1\right) g\left(i+\sum(m_j-\alpha_j)+s_n+1\right) 
\ee

Now, the idea is that with some change of variables we will be able to do all the $l$ sums. To that end change $l_n \to l_n'=l_n-l_1-\cdots-l_{n-1}$ and move the $l_n'$ sum inside the other $l$ sums and move all of them inside the $s_n$ sum. One can then show 
\be
\sum_{l_n=0}^{i+\sum (m_j-k_j) +1} \sum_{l_1+ \cdots + l_{n-1}=0}^{l_n} \sum_{s_n=0}^{\sum (\alpha_j-k_j)-l_n } & = & \sum_{s_n=0}^{\sum(\alpha_j-k_j)} \sum_{l_1+ \cdots + l_{n-1}=0}^{\sum(\alpha_j-k_j)-s_n} \sum_{l_n'=0}^{\sum(\alpha_j-k_j)-s_n-l_1-\cdots-l_{n-1}}
\ee
\noindent
using the fact that $\sum \alpha_j \le i+\sum m_j+1$. Now $l_n'$ appears in only two binomial coefficients and we can evaluate the sum to be 
\be
& & \sum_{l_n'=0}^{\sum(\alpha_j-k_j)-s_n-l_1-\cdots-l_{n-1}} (-1)^{l_n'} \binom{i+m_n+1-l_1-\cdots-l_{n-1}-\sum k_j}{l_n'} \cr
& \times &  \binom{i+m_n+1-\sum k_j-l_1-\cdots-l_{n-1}-l_n'}{\alpha_n-k_n-s_n+s_1+\cdots+s_{n-1}-l_n'} =  \delta_{0,\alpha_n-k_n-s_n+s_1+\cdots+s_{n-1}}
\ee
\noindent
where we have used the fact that $\sum(\alpha_j-k_j)-s_n-l_1-\cdots-l_{n-1} > \alpha_n-k_n-s_n+s_1+\cdots+s_{n-1}$ and hence the above sum can be evaluated using the Chu-Vandermonde identity.

Now, we attempt the other $l_j$ sums. Notice that they too appear in only two binomial coefficients. We first separate the sum as follows
\be
\sum_{l_1+ \cdots + l_{n-1}=0}^{\sum(\alpha_j-k_j)-s_n}=\sum_{l_1=0}^{\sum(\alpha_j-k_j)-s_n}\sum_{l_2=0}^{\sum(\alpha_j-k_j)-s_n-l_1} \cdots \sum_{l_{n-1}=0}^{\sum(\alpha_j-k_j)-s_n-l_1-\cdots-l_{n-2}}
\ee

We then have
\be
& & \sum_{l_{n-1}=0}^{\sum(\alpha_j-k_j)-s_n-l_1-\cdots-l_{n-2}} (-1)^{l_{n-1}} \binom{m_{n-1}+l_{n-1}-1}{l_{n-1}} \cr
& \times & \binom{m_{n-1}}{\alpha_{n-1}-k_{n-1}-l_{n-1}-s_{n-1}} = \delta_{0,\alpha_{n-1}-k_{n-1}-s_{n-1}}
\ee
\noindent
which can be summed because $\alpha_{n-1}-k_{n-1}-s_{n-1}<\sum(\alpha_j-k_j)-s_n-l_1-\cdots-l_{n-2}$ as can be verified by substituting the value of $\alpha_n$ and cancelling terms.

Similarly, one can sum each of the $l_j$ sums in precisely the same way and the verification process goes through each time by substituting the other $\alpha_j$'s. Using all these Kronecker $\delta$'s, we can substitute the $s_j$'s to yield a sum that depends only on the various $k_j$'s. Then, one notices that $\sum k_j =\sum \alpha_j - s_n$. Thus we get a factor of $g^2(i+\sum (m_j-k_j)+1)$ . Now, let us stop at this stage for a moment and evaluate the other term $Z^\dag Z |i \rangle$.
\be
Z^\dag Z |i \rangle & = & \sum_{k_1=0}^{m_1} \cdots \sum_{k_{n-1}=0}^{m_{n-1}} \sum_{k_n=0}^{i+m_n} \sum_{l_n=0}^{i+\sum (m_j-k_j) -1} \sum_{l_1+ \cdots + l_{n-1}=0}^{l_n} \sum_{r_1=0}^{m_1} \cdots \sum_{r_{n-1}=0}^{m_{n-1}} \sum_{r_n=0}^{i+m_n-l_n-1-\sum k_j} \cr
& \times & \sum_{s_n=0}^{i-l_n+\sum (m_j-k_j-r_j) } \sum_{s_1+ \cdots + s_{n-1}=0}^{s_n} (-1)^{l_n+s_n} \binom{m_1}{k_1} \cdots \binom{m_{n-1}}{k_{n-1}} \binom{i+m_n}{k_n}\cr
& \times &  \binom{m_1+l_1-1}{l_1} \cdots \binom{m_{n-1}+l_{n-1}-1}{l_{n-1}} \binom{m_1}{r_1} \cdots \binom{m_{n-1}}{r_{n-1}}\cr
& \times &   \binom{i+m_n-1-l_n-\sum k_j}{r_n} \binom{m_1+s_1-1}{s_1} \cdots \binom{m_{n-1}+s_{n-1}-1}{s_{n-1}} \cr
& \times &  \binom{i+m_n-1-l_1-\cdots-l_{n-1}-\sum k_j}{l_n-l_1-\cdots-l_{n-1}} \cr
& \times &  \binom{i+m_n-l_n-s_1-\cdots-s_{n-1}-\sum (k_j+r_j)}{s_n-s_1-\cdots-s_{n-1}} \cr
& \times & g\left(i+\sum(m_j-k_j)\right) g\left(i-l_n+\sum(m_j-k_j-r_j)\right) \cr
& \times & p_1^{k_1+l_1+r_1+s_1} \cdots p_{n-1}^{k_{n-1}+l_{n-1}+r_{n-1}+s_{n-1}} p_n^{k_n+l_n+r_n+s_n-l_1-\cdots -l_{n-1} -s_1-\cdots -s_{n-1}} \cr
& \times &  |i-l_n-s_n-\sum (k_j+r_j) \rangle
\ee

Now we follow exactly the same procedure as for the other term. We fix the exponents of the $p_j$'s and call them $\alpha_j$'s. We then change $l_n \to l_n'$ as before and change the order of the sums. We still have exactly the same sums as above because $\sum \alpha_j \leq i+\sum m_j -1$ holds unless all the $m_j$'s are zero, in which case the extended partial isometry becomes the identity operator and which we do not allow. So we have
\be
& & \sum_{l_n'=0}^{\sum(\alpha_j-k_j)-s_n-l_1-\cdots-l_{n-1}} (-1)^{l_n'} \binom{i+m_n-1-l_1-\cdots-l_{n-1}-\sum k_j}{l_n'} \cr
& \times &\binom{i+m_n-1-\sum k_j-l_1-\cdots-l_{n-1}-l_n'}{\alpha_n-k_n-s_n+s_1+\cdots+s_{n-1}-l_n'} = \delta_{0,\alpha_n-k_n-s_n+s_1+\cdots+s_{n-1}}
\ee
\noindent
exactly as before. At this point, we have almost the same sum as before except that the arguments for the $g$-functions are slightly different. The factors are $g\left(i+\sum(m_j-k_j)\right) \\ g\left(i+\sum(m_j-\alpha_j)+s_n\right)$. All the binomial coefficients are exactly the same and so, we can perform the $l_j$ sums in exactly the same way we did before.

Now, we combine both terms to get
\be
[Z,Z^\dag] |i \rangle & = & \sum_{k_1=0}^{m_1} \cdots \sum_{k_{n-1}=0}^{m_{n-1}} \sum_{k_n=0}^{i+m_n} (-1)^{\sum (\alpha_j-k_j)} \binom{m_1}{k_1} \cdots \binom{m_{n-1}}{k_{n-1}} \binom{i+m_n}{k_n}  \cr
& \times &  \binom{m_1+\alpha_1-k_1-1}{\alpha_1-k_1} \cdots \binom{m_{n-1}+\alpha_{n-1}-k_{n-1}-1}{\alpha_{n-1}-k_{n-1}} \cr
& \times & \binom{i+m_n-\alpha_1-\cdots-\alpha_{n-1}-k_n}{\alpha_n-k_n} p_1^{\alpha_1} \cdots p_n^{\alpha_n} |i - \sum \alpha_j \rangle \cr
& \times & \Big[g^2\left(i+\sum(m_j-k_j)+1\right)-g^2\left(i+\sum(m_j-k_j)\right)\Big]
\ee

Now, let $h(k_1,\ldots,k_n) = g^2\left(i+\sum(m_j-k_j)+1\right)-g^2\left(i+\sum(m_j-k_j)\right)$. We expand $h$ in a Taylor series but instead of using monomials in $k_j$'s, we use falling factorials. That is, $(k)_j=k(k-1)\cdots(k-j+1)$. The reason for doing that is that we can then evaluate each of the $k_j$ sums simply. For example,
\be
\sum_{k=0}^{m} (-1)^{k} \binom{m}{k} \binom{m+\alpha-k-1}{\alpha-k} (k)_j = \delta_{\alpha,j}
\ee

So, let $h(k_1,\ldots,k_n)=\sum_{t_1,\ldots,t_n} h_{t_1,\ldots,t_n} (k_1)_{t_1} \cdots (k_n)_{t_n}$. Taking the $t_j$ sums outside, we can do each of the $k_j$ sums. This will force $\alpha_j = t_j$ and will thus be a monomial in $p_j$. But now recall that we want $[Z,Z^\dag]=1$ and thus, we want the only non-zero term in the Taylor expansion of $h$ to by $h_{0,\ldots,0}$. Then, evaluating the sums forces all the $\alpha_j$'s to be 0 and we get $h_{0,\ldots,0} |i \rangle$ and thus, we have to force $h(k_1,\ldots,k_n)=1$.

This means $g^2\left(i+\sum(m_j-k_j)+1\right)-g^2\left(i+\sum(m_j-k_j)\right)=1$, which is a first order difference equation. For the initial conditions, it makes sense to take $g(0)=g(1)=\cdots=g(\sum m_j)=0$ because $g$ never acts on them. With this initial condition, we get a unique solution $g(N)=\sqrt{N-\sum m_j}$. This gives a unique solution for $f$, namely
\be
f(N) = \sqrt{1-\frac{m_1+\cdots+m_n}{N}}
\ee
\noindent
which completes the proof.

Just as for the 2-instanton case, notice that the proof of the extended partial isometry in the previous section and the proof of the instanton equation in this section would have gone through if we had clubbed $i$ with any of the $m_j, j=0,\ldots,n-1$ instead of $m_n$. Therefore, the same $m$-instanton is described in exactly $m$ ways. One can thus see that the moduli space of extended partial isometries is larger than the Hilbert Scheme of Points.

\section{Generalizing to Higher Dimensions}
One could also conjecture that the moduli space of extended partial isometries in higher dimensions is atleast  Hilb$_n[\mathbb{C}^m]$. We write down the extended partial isometries in any dimension for an $m$-instanton. We do not, however, either prove that they do satisfy the isometry condition or that they satisfy the instanton equation in general. The reason is that it is quite tedious, although it is straightforward. We give the proof for the 1-instanton first to highlight the salient points and point out changes from the one dimensional case.

We will work with the case of two complex dimensions for simplicity. The generalization to higher dimensions will be evident. As in the one dimensional case, we would like to start with the shift isometry. Thus, we would like $S,S^\dag$ such that
\be
S S^\dag & = & I\\
S^\dag S & = & I - \sum_{\substack{a,b > 0 \\ a+b<m}}| a, b \rangle \langle a,b|
\ee

Then postulating the same ansatz as in (\ref{ansatz}) and using the methods of \cite{Kraus:2001xt}, it is not too hard to see that the function
\be
g(N) = \sqrt{1-\frac{m(m+1)}{N(N+1)}}
\ee
\noindent
does the job. For our purpose, however, we need to know the exact form of $S,S^\dag$. One could do this is multiple ways. We label the vectors not by $|i,j \rangle$ which is the usual choice but by $|i+j,i \rangle$. Then
\be
S^\dag |i+j,i \rangle & = & |i+j+m,i \rangle \\
S|i+j,i \rangle & = & |i+j-m,i \rangle
\ee
\noindent
with the understanding that if $i+j<0$, then $|i+j,i \rangle=0$ and if $i+j<i$, then $|i+j,i \rangle= |i+j,i+j \rangle$. Then going through the exercise, we get the difference equation
\be
(N+2)g^2(N+1)-N g^2(N) = 2 \qquad g(m)=0
\ee
\noindent
for which the previous function is indeed the unique solution. As before, this corresponds to the 1-instanton at the origin. Now, we would like to generalize this to the 1-instanton at the point $(x,y)$. Again, we generalize to the notation of $T,S$ where $T$ raises and $S$ lowers and $T \neq S^\dag$. The solution is then
\be
T |i+j,i \rangle & = & \sum_{k=0}^{j+m-1} \sum_{l=0}^{i+1} \binom{j+m-1}k \binom{i+1}l x^k y^l |i+j+m-k-l,i+1-l \rangle \\
S|i+j,i \rangle & = & \sum_{k=0}^j \sum_{l=0}^i (-1)^{k+l} \binom{j}k \binom{i}l x^k y^l |i+j-m-k-l,i-1-l \rangle
\ee
\noindent
where we have taken the ideal to be of the form $(p-x)^{j+m-1} (q-y)^{i+1}$ in the polynomial ring $\mathbb{C}[p,q]$. We might have taken any other ideal of the form $(p-x)^{j+m-k} (q-y)^{i+k}$ where $k=1,\ldots,m-1$. Once again, we see the same idea that appeared in one dimension. There are multiple extended partial isometries that will yield the same instanton configuration.

Now, it can be shown that $ST=I$ because the sums split conveniently into the $j$ and $i$ factors once the powers of $x,y$ are fixed. The proof then goes through the same way as it did in one dimension.

One can construct the algorithm for the $r$-instanton in $d$ dimensions in much the same way as in the one dimensional case. Consider polynomials in $x_1,\ldots,x_d$. As a vector space this is isomorphic to the $d$-dimensional harmonic oscillator vector space with $|n_1,\ldots,n_d \rangle \sim x_1^{n_1} \cdots x_d^{n_d}$. Then the action of $T,S$ for the $r$-instanton labelled by positive integers $m_1,\ldots,m_r$ on the points $(p^{(1)}_1,\ldots,p^{(1)}_d),\cdots,(p^{(r)}_1,\ldots,p^{(r)}_d))$ can be described as follows
\be
T(x_1^{n_1} \cdots x_d^{n_d}) & = & \prod_{l=1}^d \prod_{k=1}^{r-1} (x_l+p^{(k)}_l)^{j^{(l)}_k} \prod_{l=1}^d (x_l+p^{(r)}_l)^{n_l+j^{(l)}_r} \cr
S(x_1^{n_1} \cdots x_d^{n_d}) & = & \prod_{l=1}^d \prod_{k=1}^{r-1} (x_l+p^{(k)}_l-p^{(r)}_l)^{-j^{(l)}_k} \prod_{l=1}^d x_l^{-j^{(l)}_r} (x_l-p^{(r)}_l)^{n_l}
\ee
\noindent
where $j^{(l)}_k$ are partitions of $m_k$. That is $\sum_{l=1}^d j^{(l)}_k = m_k$ and each $j^{(l)}_k > 0$. Note that this forces each of the $m_k \geq d$. Therefore one cannot construct solutions with the number of actual instantons being less than the dimension this way. This is rather strange because one can get a true partial isometry with the codimension of the ideal being 1 in any dimension whatsoever. Since, at any dimension, there are only a finite number of these cases, it is also possible that they are handled separately.

Now, using the same ansatz for $Z_i, Z^\dag_i$, we could try to solve the instanton equation for an arbitrary function. We first demonstrate how this works for the 1-instanton in two dimensions.

Now using the ansatz, (\ref{ansatz2}), we write down expressions for $Z_{1,2}$ and $Z^\dag_{1,2}$.
\be
Z_1 |i+j,i \rangle & = & \sum_{k=0}^{j+m-1} \sum_{l=0}^{i+1} \sum_{p=0}^{j+m-1-k} \sum_{q=0}^{i-l} (-1)^{p+q} x^{k+p} y^{l+q} \binom{j+m-1}k \binom{i+1}l \cr 
& \times & \binom{j+m-1-k}p \binom{i-l}q g(i+j+m-k-l) \cr
& \times & \sqrt{i-l+1} |i+j-k-l-p-q-1,i-l-q-1 \rangle\\
Z_2 |i+j,i \rangle & = & \sum_{k=0}^{j+m-1} \sum_{l=0}^{i+1} \sum_{p=0}^{j+m-k-2} \sum_{q=0}^{i-l+1} (-1)^{p+q} x^{k+p} y^{l+q} \binom{j+m-1}k \binom{i+1}l \cr 
& \times &  \binom{j+m-k-2}p \binom{i-l+1}q g(i+j+m-k-l)\cr
& \times & \sqrt{j+m-k-1} |i+j-k-l-p-q-1,i-l-q \rangle\\
Z^\dag_1 |i+j,i \rangle & = & \sum_{k=0}^{j+m-1} \sum_{l=0}^{i+1} \sum_{p=0}^{j+m-1-k} \sum_{q=0}^{i-l+2} (-1)^{p+q} x^{k+p} y^{l+q} \binom{j+m-1}k \binom{i+1}l \cr
& \times &  \binom{j+m-1-k}p \binom{i-l+2}q g(i+j+m-k-l+1)\cr
& \times & \sqrt{i-l+2} |i+j-k-l-p-q+1,i-l-q+1 \rangle\\
Z^\dag_2 |i+j,i \rangle & = & \sum_{k=0}^{j+m-1} \sum_{l=0}^{i+1} \sum_{p=0}^{j+m-k} \sum_{q=0}^{i-l+1} (-1)^{p+q} x^{k+p} y^{l+q} \binom{j+m-1}k \binom{i+1}l \cr
& \times &  \binom{j+m-k}p \binom{i-l+1}q g(i+j+m-k-l+1)\cr
& \times & \sqrt{j+m-k} |i+j-k-l-p-q+1,i-l-q \rangle
\ee

The function turns out to be entirely determined by the equation $[Z_i,Z^\dag_i]=2$ because of the nature of the ansatz as explained before. We start out by writing down the expressions for the four terms. They are as follows:
\be
& & Z_1 Z^\dag_1|i+j,i \rangle = \sum_{k=0}^{j+m-1} \sum_{l=0}^{i+1} \sum_{p=0}^{j+m-1-k} \sum_{q=0}^{i-l+2} \sum_{k2=0}^{j-k-p+m-1} \sum_{l2=0}^{i-l-q+2} \sum_{p2=0}^{j+m-1-k-p-k2} \sum_{q2=0}^{i-l-q-l2+1}   \cr
& \times & (-1)^{p+q+p2+q2}  x^{k+p+k2+p2} y^{l+q+l2+q2} \binom{j+m-1}k \binom{i+1}l \binom{j+m-1-k}p   \cr
& \times &  \binom{i-l+2}q \binom{j-k-p+m-1}{k2} \binom{i-l-q+2}{l2} \binom{j+m-1-k-p-k2}{p2}  \cr
& \times & \binom{i-l-q-l2+1}{q2} \sqrt{i-l+2} \sqrt{i-l-q-l2+2} \cr
& \times & g(i+j-k-l-p-q+m-k2-l2+1) g(i+j+m-k-l+1) \cr
& \times & |i+j-k-l-p-q-k2-l2-p2-q2,i-l-q-l2-q2 \rangle  \label{z1z1bar} \\
& & Z^\dag_1 Z_1|i+j,i \rangle = \sum_{k=0}^{j+m-1} \sum_{l=0}^{i+1} \sum_{p=0}^{j+m-1-k} \sum_{q=0}^{i-l} \sum_{k2=0}^{j-k-p+m-1} \sum_{l2=0}^{i-l-q} \sum_{p2=0}^{j+m-1-k-p-k2} \sum_{q2=0}^{i-l-q-l2+2}  \cr
& \times & (-1)^{p+q+p2+q2} x^{k+p+k2+p2} y^{l+q+l2+q2} \binom{j+m-1}k \binom{i+1}l \binom{j+m-1-k}p    \cr
& \times &  \binom{i-l}q \binom{j-k-p+m-1}{k2} \binom{i-l-q}{l2} \binom{j+m-1-k-p-k2}{p2}  \cr
& \times & \binom{i-l-q-l2+2}{q2} \sqrt{i-l+1} \sqrt{i-l-q-l2+1} \cr
& \times &    g(i+j-k-l-p-q+m-k2-l2) g(i+j+m-k-l) \cr
& \times & |i+j-k-l-p-q-k2-l2-p2-q2,i-l-q-l2-q2 \rangle \label{z1barz1} \\
& & Z_2 Z^\dag_2|i+j,i \rangle = \sum_{k=0}^{j+m-1} \sum_{l=0}^{i+1} \sum_{p=0}^{j+m-k} \sum_{q=0}^{i-l+1} \sum_{k2=0}^{j-k-p+m} \sum_{l2=0}^{i-l-q+1} \sum_{p2=0}^{j+m-1-k-p-k2} \sum_{q2=0}^{i-l-q-l2+1}  \cr
& \times & (-1)^{p+q+p2+q2} x^{k+p+k2+p2} y^{l+q+l2+q2} \binom{j+m-1}k \binom{i+1}l \binom{j+m-k}p  \cr
& \times & \binom{i-l+1}q \binom{j-k-p+m}{k2} \binom{i-l-q+1}{l2} \binom{j+m-1-k-p-k2}{p2}  \cr
& \times & \binom{i-l-q-l2+1}{q2} \sqrt{j+m-k} \sqrt{j-k-p-k2+m} \cr
& \times & g(i+j-k-l-p-q+m-k2-l2+1) g(i+j+m-k-l+1) \cr
& \times & |i+j-k-l-p-q-k2-l2-p2-q2,i-l-q-l2-q2 \rangle \label{z2z2bar} \\
& & Z^\dag_2 Z_2|i+j,i \rangle = \sum_{k=0}^{j+m-1} \sum_{l=0}^{i+1} \sum_{p=0}^{j+m-k-2} \sum_{q=0}^{i-l+1} \sum_{k2=0}^{j-k-p+m-2} \sum_{l2=0}^{i-l-q+1} \sum_{p2=0}^{j+m-1-k-p-k2} \sum_{q2=0}^{i-l-q-l2+1}   \cr
& \times & (-1)^{p+q+p2+q2} x^{k+p+k2+p2} y^{l+q+l2+q2} \binom{j+m-1}k \binom{i+1}l \binom{j+m-k-2}p   \cr
& \times & \binom{i-l+1}q \binom{j-k-p+m-2}{k2} \binom{i-l-q+1}{l2} \binom{j+m-1-k-p-k2}{p2}  \cr
& \times & \binom{i-l-q-l2+1}{q2} \sqrt{j+m-k-1} \sqrt{j-k-p-k2+m-1}\cr
& \times & g(i+j-k-l-p-q+m-k2-l2) g(i+j+m-k-l)\cr
& \times & |i+j-k-l-p-q-k2-l2-p2-q2,i-l-q-l2-q2 \rangle\label{z2barz2}
\ee
where the function $g(N)$ is to be determined.

Consider the first term (\ref{z1z1bar}). We perform all manipulations on it and claim the same go through for all the others. Define $\alpha=k+p+k2+p2, \beta=l+q+l2+q2$. Substituting for $k2,l2$, we get
\be
Z_1 Z^\dag_1|i+j,i \rangle & =  & x^\alpha y^\beta |i+j-\alpha-\beta, i - \beta \rangle \cr
& \times & \sum_{k=0}^{j+m-1} \sum_{l=0}^{i+1} \sum_{p=0}^{j+m-1-k} \sum_{q=0}^{i-l+2} (-1)^{p+q} \sqrt{i-l+2} \, g(i+j+m -k-l+1) \cr
& \times & \binom{j+m-1}k \binom{i+1}l \binom{j+m-1-k}{p} \binom{i-l+2}q \cr
& \times &  \sum_{p2=0}^{\alpha-k-p} \sum_{q2=0}^{\beta-l-q} (-1)^{p2+q2} \binom{j+m-1-k-p}{\alpha-k-p-p2} \binom{i-l-q+2}{\beta-l-q-q2}   \cr
& \times & \binom{j+m-1-\alpha+p2}{p2}  \binom{i-\beta+q2+1}{q2} \cr
& \times & \sqrt{i-\beta+q2+2} \, g(i+j+m -\alpha -\beta+p2+q2+1)
\ee

 By definition $\alpha \leq j$ and $\beta \leq i$. So we can replace the upper index in the $p$ and $q$ summations by $\alpha-k$ and $\beta -l$ respectively. Now we swap the $p,p2$ and the $q,q2$ indices. Then we can do the $p,q$ sums using the Chu-Vandermonde convolution as follows
\be
\sum_{p=0}^{\alpha -k-p2} (-1)^p \binom{j+m-1-k}p \binom{j+m-1-k-p}{\alpha -p2-k-p} = \delta_{0,\alpha-p2-k} \\
\sum_{q=0}^{\beta -l-q2} (-1)^q \binom{i+2-l}q \binom{i+2-l-q}{\beta -q2-l-q} = \delta_{0,\beta-q2-l}
\ee

So this forces $p2=\alpha-k,q2=\beta-l,k2=l2=0$. So we are left with
\be
Z_1 Z^\dag_1|i+j,i \rangle & =  & (-1)^{\alpha+\beta} x^\alpha y^\beta |i+j-\alpha-\beta, i - \beta \rangle    \cr
& \times & \sum_{k=0}^{j+m-1} \sum_{l=0}^{i+1} (-1)^{k+l} \binom{j+m-1}k \binom{i+1}l \binom{j+m-k}{\alpha-k} \binom{i-l}{\beta-l} \cr
& \times & (i-l+1) \, g^2(i+j+m-k-l+1)
\ee

Exactly the same operations go through for the other three terms also. Adding them up, we get
\be
& & \left([Z_1,Z^\dag_1] + [Z_2,Z^\dag_2] \right) |i+j,i \rangle  =  (-1)^{\alpha+\beta} x^\alpha y^\beta  |i+j-\alpha-\beta, i - \beta \rangle \binom{j+m-1}\alpha  \cr
& \times &  \binom{i+1}\beta \sum_{k=0}^{j+m-1} \sum_{l=0}^{i+1} (-1)^{k+l} \binom{\alpha}{k} \binom{\beta}{l} \left[(i+j+m-k-l+2) \times \right. \cr
&  & \left. g^2(i+j+m-k-l+1) -(i+j+m-k-l) \times g^2(i+j+m-k-l) \right]
\ee

Call the function within square brackets $h(k,l)$. We now follow the procedure in Section \ref{inst}. Expanding $h(k,l)=\sum_{r,t \geq 0} h_{r,t} (k)_r (l)_t$ in terms of falling factorials, we see that there will be polynomial contributions in $x,y$ if $h_{r,t} \neq 0$ for $r,t > 0$. Since we want the right hand side to be equal to $2 |i+j,i \rangle$, the only contribution must come from $h_{0,0}$ and it must be 2. Thus we have the same difference equation $(N+2) g^2(N+1) - N g^2(N)=2$ with the same initial condition $g(m)=0$ as for the shift isometry. This ensures that we get the same solution as before. And that this is the unique solution.

Note that the solutions here are more general than the shift isometries. That is because it is no longer true that $[Z_1,Z^\dag_1]=[Z_2,Z^\dag_2]=1$. Some extra terms naturally occur which cancel only upon the inclusion of the other term.

Notice how the proof worked. Once we defined the exponent of $x,y$, the $p$ and $q$ sums conveniently split. Both the arguments of the function $g$ as well as the square roots did not depend on either of them and so after switching summation indices, we reduce directly to a Chu-Vandermonde identity. This is exactly what happened for the 1-instanton in one dimension. In other words, the 1-instanton identity in two dimensions reduced, after fixing the exponents of the unknown quantities parametrizing the instanton locations, to two 1-instanton identities in one dimension.

One can clearly see where this is going. The $m$-instanton identity in $d$ dimensions is going to reduce to exactly $d$ $m$-instanton identities in one dimension. This is because all the binomial coefficients do not couple among different coordinates. The only quantities that link all the various coefficients are the arguments in the function $g$, the square roots and the quantities labelling the vector. When we fix the exponents of the instanton locations, the vector is fixed and this eliminates the dependence of $g$ as well as the square roots on most of the summing indices. Things just work out as if we are doing the sum for $d$ different $m$-instantons labelled by $j^{(l)}_k$ for $l=1,\ldots,d$. This causes the two square root terms to combine. Once each individual term in $Z_l Z^\dag_l$ and $Z^\dag_l Z_l$ are carried out, we can proceed to force $\sum_{l=1}^d [Z_l,Z^\dag_l] = d$. We then get a difference equation which looks like
\be
(N+d)g^2(N+1)-N g^2(N) = d
\ee 

As before, we set the initial condition $g(\sum_k m_k)=0$ because $N$ in the previous equation is always greater than or equal to $\sum_k m_k$. This forces the unique solution to be 
\be
g(N) = \sqrt{1- \frac{\prod_{l=0}^{d-1}(m_1+\cdots+m_n+l)}{\prod_{l=0}^{d-1}(N+l)}}
\ee

As explained before, there are $m-1$ extended partial isometries which can solve the 1-instanton equation with $m$ of them at point $(x,y)$ in two dimensions. we generalize this to an $m$-instanton labelled by $n_1,\ldots,n_m$ in $d$ dimensions formally.

Let $F(n,p)$ be the number of partitions of $n$ into exactly $p$ parts where each part is strictly positive. That is, $F(n,p)=0$ if $n < p$. Then the number of partial isometries for an $m$-instanton labelled by $n_1,\ldots,n_m$ in $d$ dimensions is given by $m \times \prod_{k=1}^m F(n_k,d)$.

\section{Other Theories}
Although we use the extended partial isometries in great detail to solve the noncommutative theory in \cite{Iqbal:2003ds}, it is worth wondering if the same technique can be applied to solve some other equations as well. We show that this works for at least two other theories.

\subsection{Noncommutative Yang-Mills in even dimensions}
We borrow the conventions from section 2 of \cite{Kraus:2001xt}. After complexifying the coordinates, the equation of motion is written as
\be
[Z_i,[Z^\dag_i,Z_j]] + [Z^\dag_i,[Z_i,Z_j]] = 0
\ee
Using the Jacobi identity, this can be rewritten as
\be
2 [Z^\dag_i,[Z_i,Z_j]] - [Z_j,[Z_i,Z^\dag_i]]=0
\ee

Now note that the same ansatz as for the original theory we are considering works perfectly. Both inner brackets are zero then and hence the sum is zero. Of course, there is always the question of whether there could be other kinds of solutions too with the same ansatz. This would be interesting to check.

\subsection{Abelian ASDYM on $\mathbb{R}^4$ with a self-dual background}
Consider the Abelian Seiberg-Witten monopole equations on a noncommutative deformation of $\mathbb{R}^4$. One particular limit of these equations are the Abelian Anti-Self-Dual Yang Mills (ASDYM) equations considered in Section 4.3 of \cite{Popov:2003xg}. Using their notation, we have
\be
[X_{z^1},X_{\bar{z}^{\bar{1}}}] + [X_{z^2},X_{\bar{z}^{\bar{2}}}] + \theta_{1 \bar{1}} + \theta_{2 \bar{2}} = 0, \qquad [X_{z^1},X_{z^2}] = 0
\ee

If we consider a self-dual background, $\theta_{1 \bar{1}} = \theta_{2 \bar{2}} := \theta$. Now, with a very similar ansatz, namely,
\be
X_{z^i} & = & - \frac{1}{\sqrt{\theta}} S a_i f(N) S^\dag \\
X_{\bar{z}^{\bar{i}}} & = &  - \frac{1}{\sqrt{\theta}}S f(N) a^\dag_i S^\dag \qquad i=1,2
\ee
\noindent
we are left with the exact same equations that we solved in 2 dimensions.

\subsection{A Mathematical Exercise}
Conside a system of equations similar to the one we considered in the paper. The equations of motion are
\be
[ Z^i , Z^j ] & = & [Z^\dag_i,Z^\dag_j] = 0  \cr
[ Z^i , Z^\dag_j ] & = & 0 \qquad i,j = 1,2,3
\ee

With the same ansatz, if we choose the initial condition $f(m)=1$ (though the reason for doing this is not very clear), then we get the function $f(N)=\sqrt{\frac{m}{N}}, \  \forall N\geq m$ in one dimension, which is sort of the trivial case. Unlike the case analyzed in the paper, we would get the same function in all dimensions because there is just a single commutator each time.

%\section{Comparisons}
%Note that unlike the solution in \cite{Martinec:2001hh}, we do not find any problem for values of $|p|$ smaller than the non-commutativity parameter $\theta$. Certainly the extended partial isometries do not care about the parameter. It might have been possible that the equations force the validity of the solutions of $|p|$ only above a certain threshold value. We do not see any such effect. I am not sure if that is a source of concern or not.

\section{Conclusions}
The ideas so far only scratch the surface of what appears to be a very rich field of study in computational combinatorics. We have proved the moduli space of extended partial isometries is atleast the Hilbert Scheme of Points and show that it is definitely much larger. Because of the noncommutative ABS construction, the space of extended partial isometries contains as a subset the Hilbert Scheme of true partial isometries. It would be very interesting to study how this subspace is embedded within the larger space of extended partial isometries.

Further, we showed that all these extended partial isometries yield valid solutions to a particular instanton equation with the same function. It would be interesting to see if this goes through for instanton equations for other noncommutative gauge theories as well.

If that is not enough, there is sufficient cause to believe that these statements go through in higher dimensions too. One would like to extend both the above-mentioned ideas to higher dimensions.

\section{Acknowledgements}
The authors would like to thank D.Zeilberger, J.Lebowitz and E.Speer for useful discussions and G.Moore for correspondence. A.A. would like to thank D.-E. Diaconescu for discussions leading to the project and to acknowledge support from the Department of Physics, Rutgers University.

\appendix

\section{Similarity Transformation}
We present a proof of the claim made in Sec \ref{2i1o}. Namely, that the operators in the solution with $n=0$ are similar to the operators in the shift isometry, i.e. similarity between $T|i \rangle =\sum_{k=0}^i\binom{i}kp^k|i+m-k \rangle $ and $T'|i \rangle =|i+m \rangle$. To this end, we switch to matrix representations and write the infinite dimensional mappings $T$ and $T'$ respectively as the infinite dimensional matrices, $T_{jk}:=\binom{k}{j-m}p^{k-j+m}$ and $T_{ij}':=\binom{0}{j-i+m}; i,j,k\geq 0.$  Then,

\noindent
\bf Claim: \rm $T$ and $T'$ are similar, i.e. $\exists$ ${}V$ such that 
$VTV^{-1}=T'$.
\\
\noindent
\bf Proof: \rm
Let $\lfloor x\rfloor$ denote \it the greatest integer \rm function. Define the infinite matrix $U$, ``Generalized'' Stirling Numbers of the 2nd kind, recursively as follows:
$U_{ij}=U_{i-1,j-1}+\lfloor\frac{j}m\rfloor U_{i-1,j}$, with initial conditions
$U_{mm}=1, U_{i,m-1}=U_{m-1,j}=0$ for $i,j\geq m$. 
Then, the inverse matrix $U^{-1}$, ``Generalized'' Stirling numbers of the 1st kind, will take the form
$U_{ij}^{-1}=U_{i-1,j-1}^{-1}+\lfloor\frac{i}m\rfloor U_{i-1,j}^{-1}$, with initial conditions
$U_{mm}^{-1}=1, U_{i,m-1}^{-1}=U_{m-1,j}^{-1}=0$ for $i,j\geq m$.(Note: for $m=1$, we do indeed get the standard Stirling numbers of the 2nd kind $U_{ij}=S(i,j)=\frac1{j!}\sum_{n=0}^j(-1)^n\binom{j}n(j-n)^i$ and their counterparts, Stirling numbers of the 1st kind, $U_{ij}^{-1}=s(i,j)$.) 
We now augment the identity matrix $I_m$, and transpose $U$, to form block matrices
\be
&V_{ij}:=\begin{cases}
\delta_{ij}\qquad\text{if $0\leq i,j\leq m-1$}\\
(-p)^{j-i}U_{j-m,i-m}\qquad\text{if $i,j\geq m$} \\ 0\qquad\text{otherwise}\end{cases}\\
&V_{ij}^{-1}:=\begin{cases}
\delta_{ij}\qquad\text{if $0\leq i,j\leq m-1$}\\
(-p)^{j-i}U_{j-m,i-m}^{-1}\qquad\text{if $i,j\geq m$}\\
0\qquad\text{otherwise.}\end{cases}
\ee

Interestingly, $V$ satisfies the recursive formula: $V_{ij}=V_{i-1,j-1}-p\lfloor\frac{i}m\rfloor V_{i,j-1}$, with initial conditions
$V_{0,0}=1, V_{i,0}=V_{0,j}=0$ for $i,j\geq 1$. A similar relation is valid for $V^{-1}$. The corresponding generating functions for the $i$th row of $V$, and the $j$th column of $V^{-1}$, can be written respectively as
\be
& \displaystyle \frac{x^{i+m}}{(1+\lfloor\frac{i}m\rfloor px)^{1+(i \mod m)}} \prod_{k=1}^{\lfloor\frac{i}m\rfloor -1}\frac1{(1+kpx)^{m}} \\
& \displaystyle x^{j+m}\left(1+\lfloor\frac{j}m\rfloor px\right)^{1+(j \mod m)} \, \prod_{k=1}^{\lfloor\frac{j}m\rfloor -1}(1+kpx)^{m}
\ee

\noindent
We will show that for each fixed $i$ and all $k$, we have
\be
V_{ij}T_{jk}=V_{jk}\binom{k}{j-m}p^{k-j+m} = T_{ij}'V_{jk}=\delta_{i-m,j}V_{jk}=V_{i-m,k}
\label{ind}
\ee
\noindent
where Einstein summation over repeated indices is in effect. We proceed by double induction on $i,k\geq 0$. For $i=0$ (in fact, any $i<m$)and $\forall k$, trivially $V_{ij}T_{jk}=\delta_{i,j}\binom{k}{j-m}p^{k-j+m}=\binom{k}{i-m}p^{k-i+m}=0$, and also $V_{i-m,k}=0$.
Now, assume (\ref{ind}) holds for all indices less than $i$ ($\forall k$). We prove for $i$ $(\forall k)$. Next, we do a sub-induction on $k$. If $k=0$, then $V_{i,j}T_{j,0}=V_{i,m}=\delta_{i,m}$ and $V_{i-m,0}=\delta_{i,m}$. Hence, assume the statement is valid for this fixed $i$ and all indices less than $k$. Then, combining these two induction assumptions we are lead to
\be
V_{ij}T_{jk}& = & V_{ij}\left\{\binom{k-1}{j-1-m}+\binom{k-1}{j-m}\right\}p^{k-j+m} \\
& = & V_{ij}\binom{k-1}{j-1-m}p^{k-j+m}+pV_{i-m,k-1}\\
& = & \left\{V_{i-1,j-1}-p\lfloor\frac{i}m\rfloor V_{i,j-1}\right\} \binom{k-1}{j-1-m}p^{k-j+m}+pV_{i-1,k-m}\\
& = & V_{i-1-m,k-1}-p\lfloor\frac{i}m\rfloor V_{i-m,k-1}+pV_{i-m,k-1} \\
& = & V_{i-1-m,k-1}-p\lfloor\frac{i-m}m\rfloor V_{i-m,k-1}\\
& = & V_{i-m,k}
\ee
This indeed completes the proof.

\end{document}